\documentclass[aps,prb,reprint,twocolumn,superscriptaddress]{revtex4-2}
\usepackage{times} 
\usepackage{mathptmx} 
\usepackage{amsmath} 
\usepackage{braket} 
\usepackage{graphicx} 
\usepackage{stackengine} 
\usepackage{appendix} 
\usepackage[colorlinks=true,
linkcolor=red, 
anchorcolor=blue,
citecolor=blue,
urlcolor=blue
]{hyperref}
\newcommand{\figref}[2]{Fig.~\hyperref[#1]{\ref*{#1}(#2)}}
\begin{document}
\setlength{\parskip}{0pt}  
\renewcommand{\baselinestretch}{1.2}  

\title{Violation of Luttinger's theorem in one-dimensional interacting fermions}
\author{Meng Gao}
\author{Yin Zhong}
\email{zhongy@lzu.edu.cn}

\affiliation{Key Laboratory of Quantum Theory and Applications of MoE \& School of Physical Science and Technology, Lanzhou University, Lanzhou 730000, People’s Republic of China}
\affiliation{Lanzhou Center for Theoretical Physics, Key Laboratory of Theoretical Physics of Gansu Province, Lanzhou University, Lanzhou 730000, People’s Republic of China}

\date{\today}

\begin{abstract}
Using the density matrix renormalization group method, we systematically investigate the evolution of the Luttinger integral in the one-dimensional generalized \( t\text{-}V \) model as a function of filling and interaction strength, and identify three representative phases. In the weak-coupling regime, the zero-frequency Green's function exhibits a branch-cut structure at the Fermi momentum, and the Luttinger integral accurately reflects the particle density, indicating that the Luttinger theorem holds. As the interaction increases, the spectral weight near the Fermi momentum is gradually suppressed. Interestingly, in the strong coupling regime near half-filling, this singularity is progressively destroyed, accompanied by the emergence of momentum-space zeros in the real part of the Green’s function, leading to a novel non-Fermi liquid metallic phase beyond the classic Luttinger liquid paradigm, where the Luttinger surface is no longer defined by a single singularity. While finite spectral weight remains at the original Fermi momentum, the singularity gradually diminishes. Meanwhile, zeros with negligible spectral weight appear away from this momentum, significantly affecting the integral. At exact half-filling, a single-particle gap opens, and the Green's function becomes nearly vanishing across the entire momentum space, indicating the complete suppression of low-energy electronic states consistent with the nature of an insulating charge-density-wave phase. These results suggest that the breakdown of the Luttinger theorem is not triggered by a single mechanism, but rather results from the interplay between interaction-driven evolution of excitation modes and the breaking of particle-hole symmetry, ultimately leading to a continuous reconstruction of the generalized Fermi surface from topologically protected to correlation-driven.
\end{abstract}

\maketitle
\section{Introduction}\label{Introduction}

Since its proposal by Luttinger and Ward~\cite{LuttingerWard1960}, the Luttinger theorem has become a cornerstone in the theoretical understanding of the Fermi liquid (FL) behavior. Within the framework of FL theory~\cite{1965723}, the Luttinger theorem~\cite{Luttinger1960} asserts that the volume enclosed by the Fermi surface is determined solely by the fermion density, regardless of interaction strength. This result is crucial for the conventional FL, as it implies that while interactions renormalize quasiparticle properties such as effective mass and Fermi velocity, the topology of the Fermi surface remains invariant at low temperatures.

In one-dimensional (1D) systems, however, the FL theory generally breaks down due to enhanced quantum fluctuations and reduced phase space. Instead, many 1D systems exhibit Luttinger liquid (LL) behavior~\cite{JVoit_1995,10.1093/acprof:oso/9780198525004.001.0001,KSchönhammer_2002}. In the LL, low-energy excitations are no longer described as individual quasiparticles but rather as bosonic collective modes that characterize charge and spin density fluctuations. This leads to hallmark features such as spin-charge separation (in spinful systems) and non-FL correlations. Although the LL is inherently non-FL, it nonetheless obeys a generalized Luttinger theorem~\cite{PhysRevLett.79.1110}:
\begin{equation}
	I_L = \int \frac{d^d k}{(2\pi)^d} \, \theta\left(\operatorname{Re} G(k, \omega=0)\right) = \int_{\operatorname{Re} G(k, 0) > 0} \frac{d^d k}{(2\pi)^d}, 
\end{equation}
where \( \theta \) is the step function, and \( I_L \) receives contributions from momentum states where the real part of the zero-frequency Green's function \( \operatorname{Re} G(k, \omega=0) \) is positive. For both the FL and LL, this integral yields the particle density: \( I_L = n = N_f / N \), where \( N_f \) is the total number of fermions and \( N \) is the number of lattice sites.

Recent studies have shown that the Luttinger theorem is not universally valid~\cite{PhysRevB.106.195117, PhysRevB.78.153103, PhysRevB.111.045107, PhysRevB.104.235122, SSJD00000216896}, particularly in systems lacking particle-hole (p-h) symmetry. Notable examples include the Falicov-Kimball model in the strong-coupling limit~\cite{PhysRevB.106.195117}, the generalized $t$-$V$ model~\cite{PhysRevB.78.153103}, the $t$-$J$ model~\cite{PhysRevB.111.045107}, and the Hubbard model~\cite{PhysRevB.111.045107,PhysRevB.104.235122}. In these systems, strong correlations can lead to deviations of the Luttinger integral $I_L$ from the particle density~\cite{PhysRevB.104.235122}. This indicates not only a reconstruction of the Fermi surface but also a fundamental change in low-energy excitations. For instance, Ref.~\cite{PhysRevB.78.153103} demonstrates a breakdown of the Luttinger theorem in the insulating phase of the half-filled generalized $t$-$V$ model across a metal-insulator transition. Ref.~\cite{PhysRevB.111.045107} identifies a ``Luttinger-breaking'' regime near half-filling in the two-dimensional Fermi-Hubbard model in the absence of long-range order. However, exact diagonalization (ED) suffers from the exponential growth of the Hilbert space, while determinant quantum Monte Carlo (DQMC) is hindered by the sign problem~\cite{PAN2024879}, which restricts both the accessible parameter space and the attainable numerical accuracy.

Strongly correlated low-dimensional systems exhibit rich dynamical behavior that goes beyond the Landau paradigm. Key observables such as single-particle spectra, momentum distributions, and correlation functions can be probed via cold-atom experiments~\cite{RevModPhys.80.885}, X-ray scattering~\cite{doi:10.1126/sciadv.adt7770}, and angle-resolved photoemission spectroscopy (ARPES)~\cite{RevModPhys.75.473}. These techniques serve as powerful tools for investigating unconventional metallic states, metal-insulator transitions~\cite{RevModPhys.78.17}, and low-energy excitations. Although LL theory has successfully described many 1D metallic systems, it is based on several strong assumptions, particularly the linearization of the dispersion relation and the expansion of two-body interactions for small momentum transfers near $k_F$. In realistic strongly correlated systems, these approximations may break down, as symmetry breaking and renormalization of excitation modes due to strong interactions can lead to significant deviations from conventional LL behavior. To date, no rigorous mathematical proof has settled this question. It is thus essential to explore whether 1D interacting fermion systems can host metallic states that retain conventional metallic behavior while violating the key assumptions of LL theory. The 1D generalized \(t\)-\(V\) model, with tunable interactions and adjustable p-h asymmetry, provides a natural platform for this investigation.

To bridge this gap, we employ the density matrix renormalization group (DMRG) method on a periodic chain with \( N = 102 \) sites to construct high-resolution (\( \Delta k \approx 0.02\pi \)) momentum-space Green's functions. This approach enables us to accurately capture critical features and probe the largely unexplored intermediate regime near half-filling. By tracking the evolution of dynamical observables, we identify three distinct correlation regimes:

\begin{enumerate}
	\renewcommand{\theenumi}{\roman{enumi}}
	\item In the weak-coupling limit, the system exhibits LL behavior characterized by power-law singularities. In this regime, the generalized Fermi surface is identified through nonanalyticities in $\operatorname{Re} G(k, 0)$, consistent with a well-defined Fermi momentum $k_F = n\pi$.
	\item With increasing interaction and filling, the topology of the generalized Fermi surface evolves continuously, marked by the evolution of singularities and zeros in \( \operatorname{Re} G(k, 0) \).
	\item In the half-filled strong-coupling regime, low-energy excitations are suppressed, indicating a transition to an insulating phase.
\end{enumerate}

In case~(ii), we identify a metallic phase beyond the LL paradigm, which we term a non-Luttinger liquid (NLL). It arises from the interplay of interaction-driven reconstruction of excitation modes and breaking of p-h symmetry, leading to a topological reorganization of the Fermi points. In the strongly interacting regime near half filling, a new type of 1D metallic state emerges. Its momentum distribution is smooth around $k_F$, and the spectral properties deviate significantly from those of the LL. A finite spectral weight persists at $k_F$, while the characteristic branch-cut singularities are gradually destroyed. Zeros of $\operatorname{Re} G(k,0)$ appear away from $k_F$, which, in the absence of topological protection, strongly affect $I_L$. This phase combines pronounced strong-coupling features with a reconstructed topological structure, revealing a class of non-Luttinger 1D correlated metals that, to our knowledge, has not been systematically explored in the literature. Identifying and characterizing this NLL phase constitutes a central part of the present work and will be discussed in detail below.

The remainder of this paper is organized as follows. In Sec.~\ref{Model and Method}, we introduce the model and numerical methods. Sec.~\ref{Numerical Results} presents the phase diagram and key results, which serve as the basis for our subsequent analysis. In Sec.~\ref{Deviation Analysis}, we examine the mechanism underlying the breakdown of the Luttinger theorem. Finally, Sec.~\ref{Summary and Discussion} concludes the paper with a summary and outlook.

\section{Model and Method}\label{Model and Method}

\begin{figure}[htbp]
	\centering
	\begin{minipage}{0.25\textwidth}
		\centering
		\includegraphics[width=1\linewidth, trim=150 180 230 145, clip]{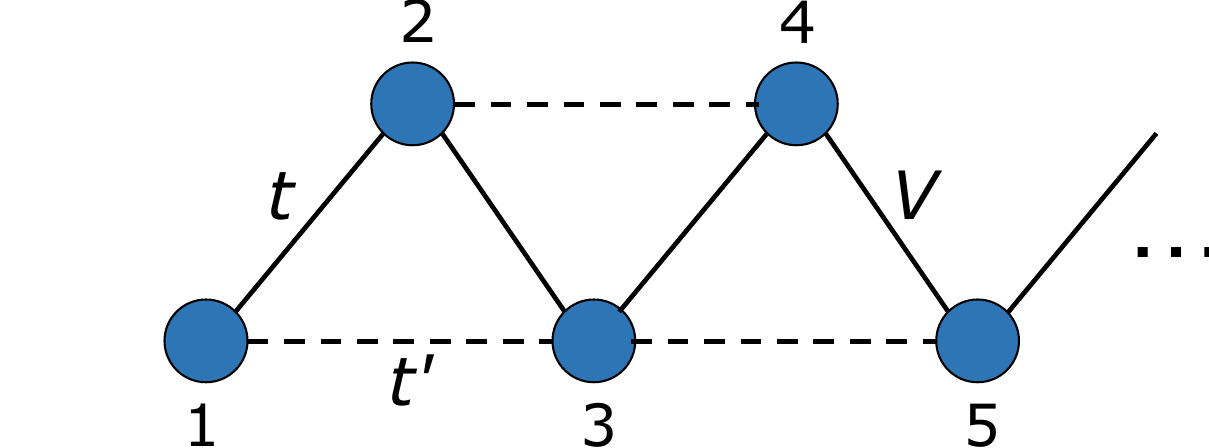}
		\par\footnotesize{(a)} 
	\end{minipage}
	\hfill
	\begin{minipage}{0.5\textwidth}
		\centering
		\includegraphics[width=0.82\linewidth, trim=80 7 60 35,clip]{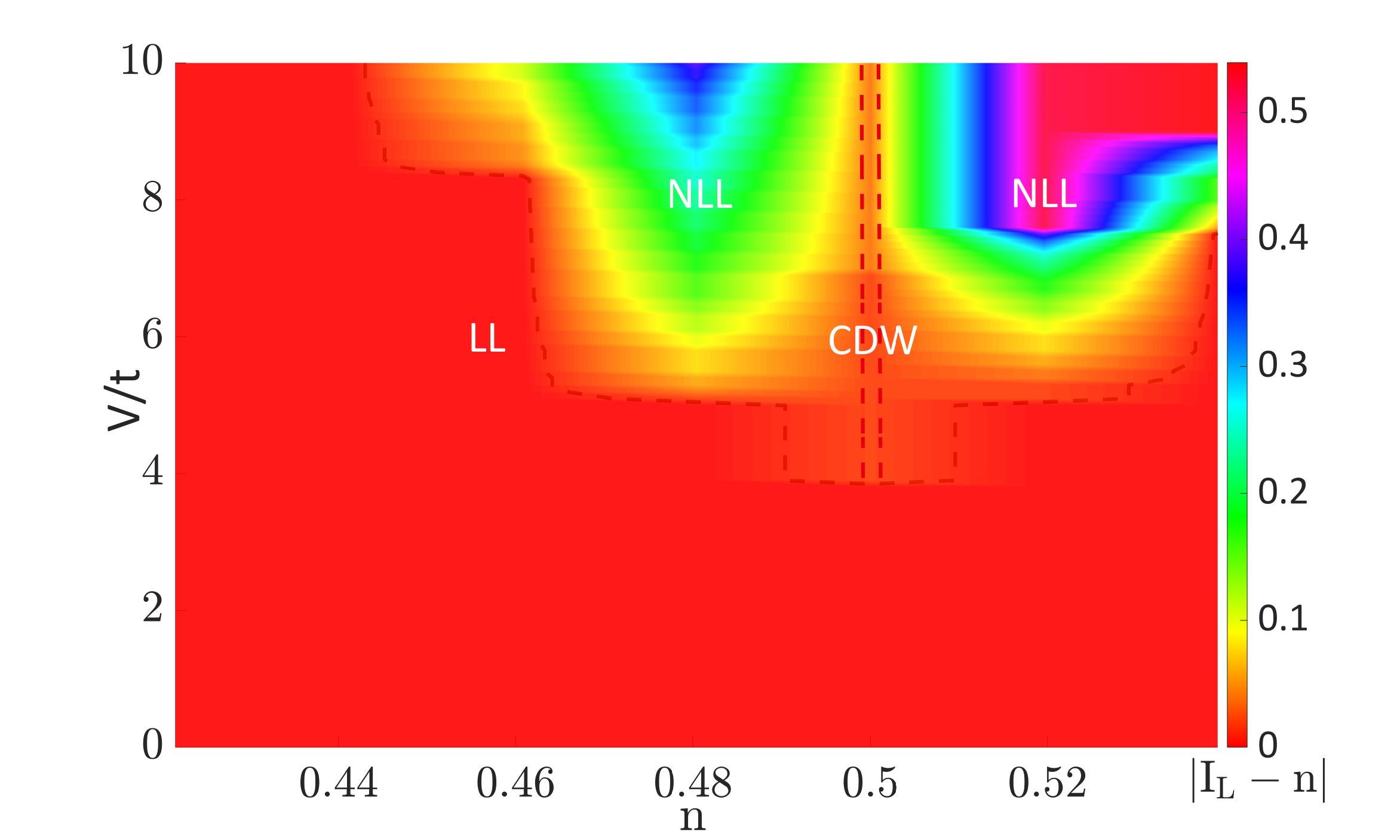}
		\par\footnotesize{(b)} 
	\end{minipage}
	\begin{minipage}{0.5\textwidth}
		\centering
		\includegraphics[width=0.95\linewidth, trim=8 12 0 0,clip]{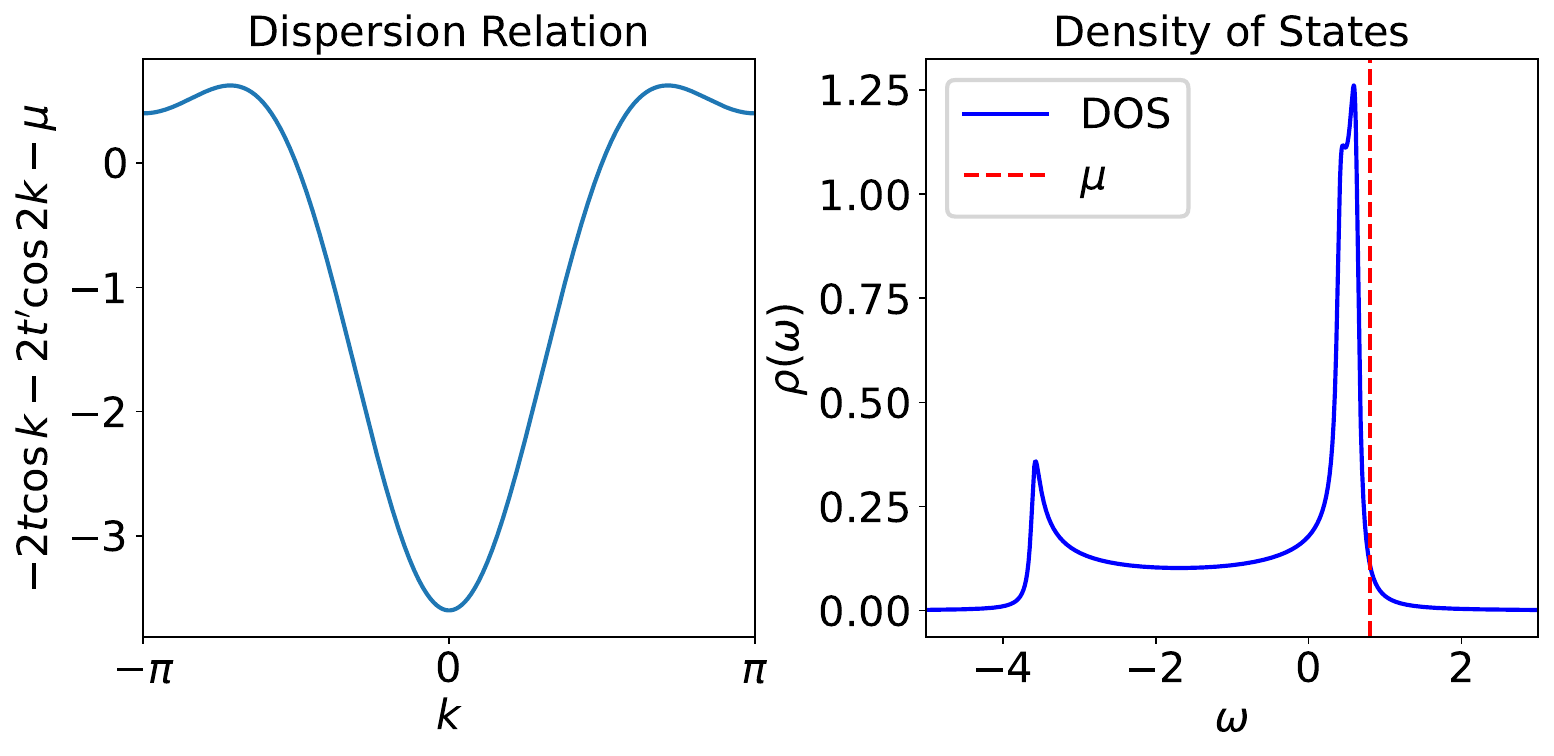}
		\par\footnotesize{(c)} 
	\end{minipage}
	\caption{(a) Schematic illustration of the Hamiltonian for the \( t \)-\( t' \)-\( V \) model.  
		(b) Zero-temperature phase diagram of the \( t \)-\( t' \)-\( V \) model in the \( n \)-\( V \) plane, with the nearest-neighbor hopping set as the energy unit (\( t = 1 \)).  
		At low filling or weak interaction, the system resides in a Luttinger liquid (LL) phase.  
		As the density \( n \) and interaction \( V/t \) increase beyond a critical threshold, the system enters a non-Luttinger liquid (NLL) phase.  
		At half-filling, an interaction-driven metal-insulator transition drives the system into a charge-density-wave (CDW) phase.  
		Beyond half-filling, the system evolves back into the NLL phase.
		(c)Non-interacting dispersion relation and the corresponding density of states (DOS) at half-filling, under the same parameters as used in (b).
		The inclusion of $t'$ shifts the position of the band maxima in the Brillouin zone and results in a clearly asymmetric DOS with respect to zero energy.}
	\label{fig:phase_diagram}
\end{figure}

To investigate 1D strongly correlated fermionic systems, LL theory provides an effective low-energy description for gapless metallic phases. However, this framework has known limitations, particularly in capturing critical behavior beyond the LL paradigm in the strong-coupling regime~\cite{PhysRevB.73.165104, PhysRevB.86.155156}. To explore such regimes, we consider a minimal but versatile model: the generalized~$t$-$V$ model (also known as the $t$-$t'$-$V$ model). This model exhibits a wide range of low-energy behaviors through tunable parameters, encompassing the free fermion gas, LL phases, and exotic strongly correlated regimes. It offers a well-controlled setting for studying non-FL behavior, highlighting its distinction from conventional FL properties and examining the applicability or potential breakdown of the Luttinger~theorem. The Hamiltonian is given by
\begin{equation}
	\hat{H} = -t \sum_i \left( \hat{c}_{i+1}^\dagger \hat{c}_i + \mathrm{H.c.} \right)
	- t' \sum_i \left( \hat{c}_{i+2}^\dagger \hat{c}_i + \mathrm{H.c.} \right)
	+ V \sum_i \hat{n}_i \hat{n}_{i+1},
\end{equation}
where $t$ and $t'$ are the nearest-neighbor and next-nearest-neighbor hopping amplitudes, respectively, and $V$ denotes the repulsive interaction between neighboring fermions (see Fig.~\ref{fig:phase_diagram}\hyperref[fig:phase_diagram]{(a)}).

To access properties in the thermodynamic limit, we employ the DMRG method~\cite{SCHOLLWOCK201196} to compute the zero-temperature single-particle Green's function:
\begin{equation}\label{green}
	G(k,\omega) = -i \int_0^{\infty} dt\, e^{i(\omega + \mu)t} \langle 0 | \{ \hat{c}_k^\dagger,\, \hat{c}_k(t) \}_+ |0 \rangle, 
\end{equation}
where $\mu$ is the chemical potential, defined via the ground-state energies as
\begin{equation}
	\mu(N) = \frac{E_0^{N_f + 1} - E_0^{N_f - 1}}{2}.
\end{equation} 

While DMRG is more naturally suited to open boundary conditions (OBC), studying finite-size systems with PBC is desirable for more direct extrapolation to the thermodynamic limit~\cite{RevModPhys.77.259}. By exploiting $U(1)$ symmetry and selecting appropriate parameters, matrix product state (MPS) representations can efficiently handle PBC systems with up to $\sim$200 lattice sites~\cite{PhysRevLett.93.227205}. In our calculations, we consider a system of size $N = 102$ to balance computational efficiency and accuracy, and systematically explore different fillings and interaction strengths.

Using the Lehmann representation, Eq.~\eqref{green} can be recast as
\begin{align}
	G(k,\omega) &= \langle 0 | \hat{c}_k^\dagger \frac{1}{\omega + \mu + E_N^0 - \hat{H}} \hat{c}_k |0 \rangle \notag \\ &\quad + \langle 0 | \hat{c}_k \frac{1}{\omega + \mu - E_N^0 + \hat{H}} \hat{c}_k^\dagger |0 \rangle. \label{eq:green}
\end{align}
where $E_N^0$ is the ground-state energy for particle number $N_f$ obtained via DMRG. The momentum-resolved creation and annihilation operators $\hat{c}_k^\dagger$ and $\hat{c}_k$ are constructed from the real-space operators via Fourier transformation:
\begin{equation}
	\hat{c}_k = \frac{1}{\sqrt{N}} \sum_{j=1}^{N} e^{-ikx_j} \hat{c}_j.
\end{equation}

In practical numerical calculations, the Hamiltonian is typically represented as a matrix product operator (MPO), and the wavefunction as a MPS, thereby avoiding the exponential complexity of the full Hilbert space. In evaluating Eq.~\eqref{eq:green}, one encounters operations involving $(z - H)^{-1}$. Since the inverse cannot be constructed explicitly, we adopt the correction vector method~\cite{PhysRevE.94.053308}, which reformulates the problem as the solution of the following linear equation:
\begin{equation}
	(z - H)|\psi\rangle = |\phi\rangle\,.
\end{equation}
This equation is efficiently solved within the tensor network framework using Krylov subspace methods, without the need to construct or store the full Hamiltonian matrix.

\section{Numerical Results}\label{Numerical Results}

For the \( t \)-\( t' \)-\( V \) model, at half-filling, the system opens a gap and enters a charge-density-wave (CDW) phase as the interaction strength increases. At other fillings, extrapolation results indicate that the system remains metallic. For example, at a fermion density \( n \approx 0.48 \), the system remains gapless even at strong coupling \( V/t = 10 \).

In Fig.~\ref{fig:phase_diagram}, we consider systems with 102 sites and \( t'/t = 0.4 \), varying the fermion density \( n \in (0.42, 0.54) \) and interaction strength \( V/t \in [0, 10] \). We study the deviation of \( I_L \) from the fermion density \( n \). At low fillings (\( n < 0.45 \)) with any finite interactions, the system behaves as an LL, where low-energy excitations are dominated by bosonized collective modes, the Fermi point remains fixed, and \( I_L \) coincides with \( n \). As \( n \) increases and the interaction strength surpasses a critical value, \( I_L \) starts to deviate from the expected value tied to the fermion density, with the deviation growing as interactions strengthen. This indicates a transition into a NLL phase, where the low-energy behavior is no longer described by LL theory and instead exhibits features characteristic of a non-FL metal. When the density approaches half-filling, the system undergoes a transition into a fully gapped CDW state once the interaction strength exceeds the critical value \( V_c = 2t \). In this phase, low-energy excitations are strongly suppressed and \( I_L \) correspondingly shifts.

\begin{figure}[htbp]
	\centering
	\includegraphics[width=1.0\linewidth]{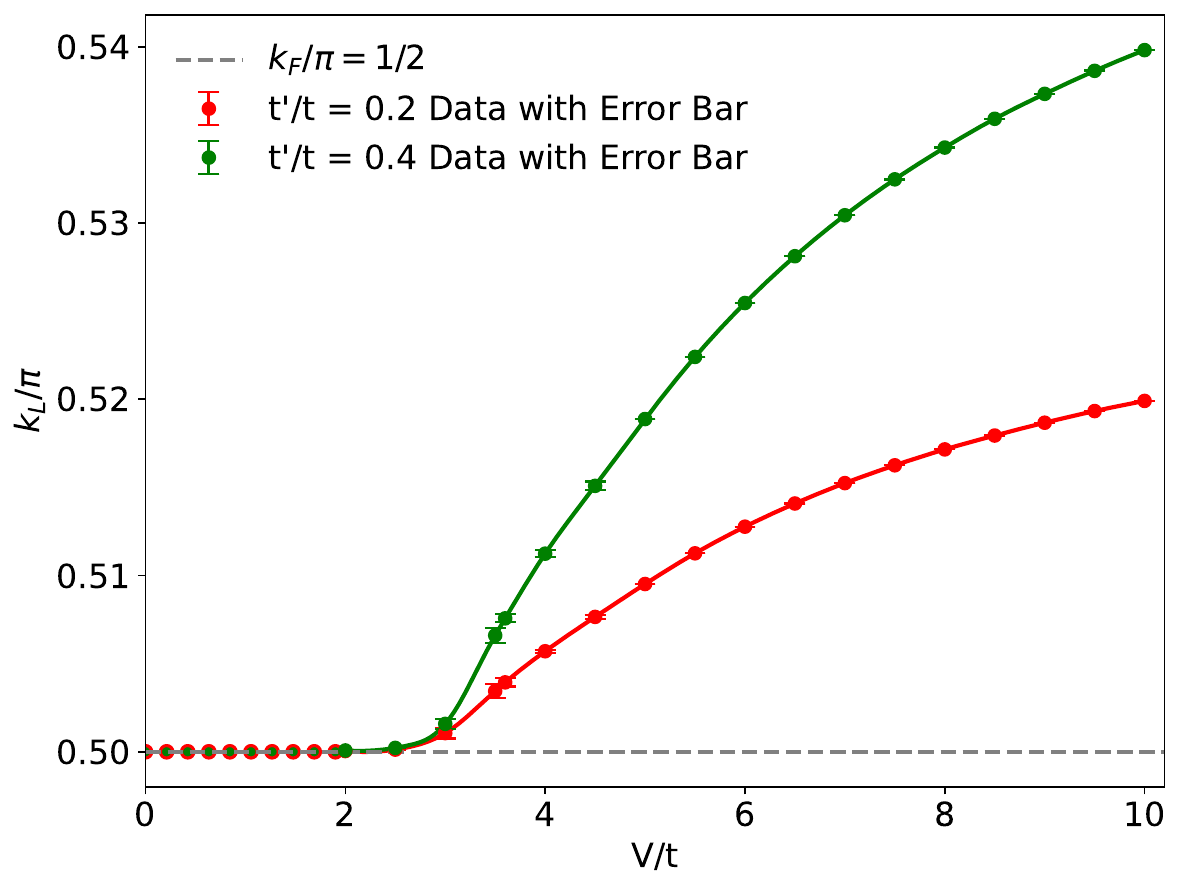} 
	\caption{Evolution of the Luttinger momentum \( k_L \) with \( V/t \) for different next-nearest-neighbor hopping strengths \( t'/t \).}
	\label{Luttinger_Momentum}
\end{figure}

It should be emphasized that the phase boundaries shown in Fig.~\ref{fig:phase_diagram} reflect crossover behavior in a finite-size system and do not correspond to sharply defined phase transitions in the thermodynamic limit. Nevertheless, the qualitative LL–NLL–CDW evolution revealed in this diagram agrees well with the scaling analyses discussed below and thus provides important physical insights. We find that under strong coupling at half-filling, the system does exhibit deviations from the Luttinger theorem. However, these deviations are largely confined to the region near half-filling (\( |n - 0.5| < 0.03 \)), especially on the over-half-filled side, where the electronic behavior becomes significantly more intricate. This suggests that the emergence of NLL features is not governed solely by electron-electron interactions. Further calculations at different values of \( t'/t \) reveal a power-law relationship between the degree of deviation and the p-h asymmetry of the system, as clearly demonstrated in Fig.~\ref{Luttinger_Momentum}.

To verify the reliability of our numerical results, we compared DMRG data with ED results for small systems, finding excellent agreement within the comparable parameter regimes (see \hyperref[app:benchmark]{Appendix}). Our findings are consistent with the ED extrapolations reported in Ref.~\cite{PhysRevB.78.153103} for systems with \( N=14 \)-\(30 \) sites. The main improvement in this work is the use of larger systems (\( N=82, 86, 90, 94, 98, 102 \)), which reduce finite-size effects and significantly improve computational accuracy, particularly addressing previous limitations in the intermediate coupling regime $ 2 < V/t < 4 $ where momentum resolution was constrained.

We also computed \( G(k, \omega=0) \) at half-filling for \( t'/t = 0.2, 0.4 \), extrapolating the Luttinger momentum \( k_L \)~\cite{PhysRevB.68.085113} to the thermodynamic limit, as shown in Fig.~\ref{Luttinger_Momentum}. For the half-filled system with \( t'=0 \), due to the exact p-h symmetry, the real part of the Green’s function satisfies the relation
\begin{equation}\label{antisymmetry}
	\mathrm{Re}\, G(k-k_F, 0) = -\mathrm{Re}\, G(k_F-k, 0),
\end{equation}
which ensures that regions with \( \mathrm{Re}\, G(k, 0) > 0 \) and \( < 0 \) are strictly antisymmetric in momentum space about \( k_F \), leading to \( I_L = n \). This result holds regardless of interaction strength and directly reflects the p-h symmetry of the model~\cite{PhysRevB.96.085124}. When a next-nearest-neighbor hopping \( t' \neq 0 \) is introduced, the exact p-h symmetry is broken, and Eq.~\eqref{antisymmetry} no longer holds strictly. Nevertheless, in the low-energy LL regime, emergent p-h and Lorentz symmetries appear, so that Eq.~\eqref{antisymmetry} remains approximately valid. The main source of uncertainty in the data points comes from the finite-size extrapolation process. The fitting curves for \( t'/t = 0.2, 0.4 \) clearly show a power-law correlation between the deviation of \( I_L \) and the degree of p-h asymmetry.

Fig.~\ref{Luttinger_Momentum} presents the evolution of the Luttinger momentum \( k_L \) with \( V/t \) for different hopping ratios \( t'/t = 0.2, 0.4 \). Here, \( k_L \) denotes the momentum where \( \operatorname{Re} G(k, 0) \) changes sign via infinity or zero.

\section{Deviation Analysis}\label{Deviation Analysis}
In FL, many-body interactions reduce the quasiparticle weight $Z$ and redistribute part of the spectral weight into an incoherent background at higher energies. The momentum distribution function (MDF)
\begin{equation}
	n(k) = \langle \hat{c}_k^\dagger \hat{c}_k \rangle,
\end{equation}
exhibits a discontinuity at $k_F$, whose magnitude is given by $Z$. As the interaction strength increases, this discontinuity diminishes, reflecting the decay of single-particle excitations. In 1D interacting systems, however, the step-like behavior is typically absent and replaced by power-law singularities and other NFL features. As such, the MDF serves as a key probe of low-energy physics, providing a means to distinguish FL, LL and other NFL states.

\begin{figure*}[htbp]
	\centering
	\includegraphics[width=1\linewidth]{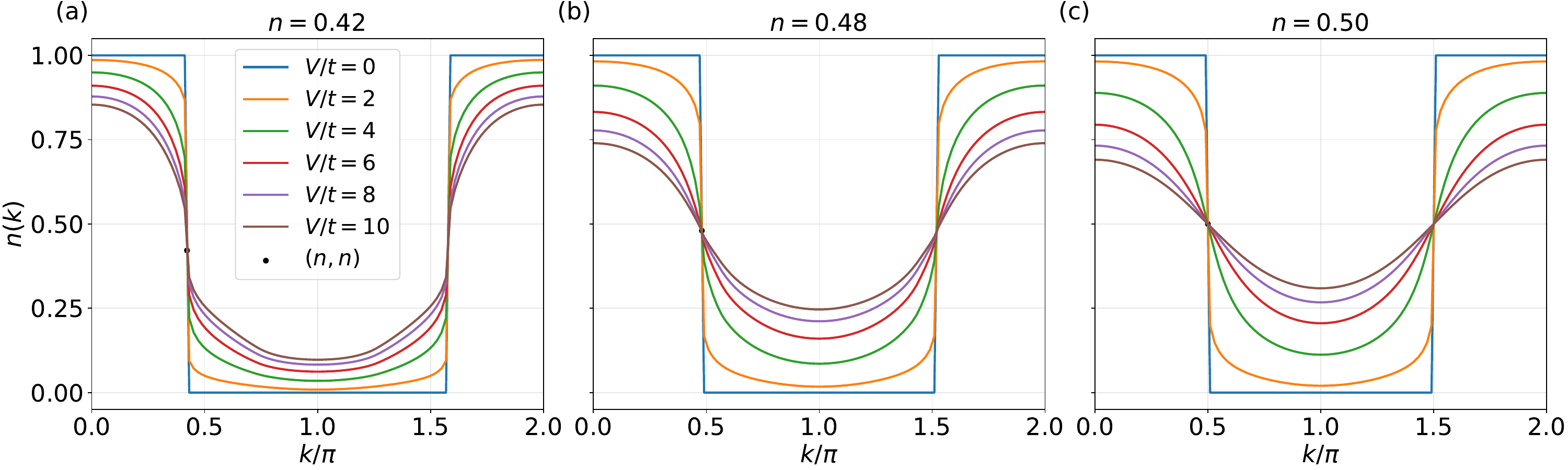}
	\caption{Evolution of the momentum distribution function $n(k)$ with interaction strength $V/t$ at $t'/t = 0.4$ for particle densities $n = 0.42$, $0.48$, and $0.5$, representing LL, NLL, and CDW regimes, respectively. 
		(a) Far from half-filling, $n(k)$ exhibits a power-law singularity near $k_F$, characterized by a Luttinger parameter $K$ that decreases with increasing $V/t$.
		(b)(c) For stronger interactions beyond a critical $V_c$, the singularity near $k_F$ smoothens out, indicating a transition to an NFL regime without a sharp Fermi surface or, at half-filling, to a CDW phase with long-range charge order.}
	\label{fig:mdf}
\end{figure*}

As shown in Fig.~\ref{fig:mdf}, when $V/t = 0$, the MDF $n(k)$ exhibits a sharp Fermi step with quasiparticle weight $Z = 1$. In the weak-coupling regime, the system displays power-law behavior characteristic of LL physics. For systems far from half-filling [\figref{fig:mdf}{a}], $n(k)$ retains its LL-type nonanalyticity near $k_F$ even at relatively strong interactions, with the Fermi point remaining fixed. In this regime, the system is well described by the Tomonaga-Luttinger model, where interactions merely renormalize the excitation velocities and correlation exponents.

In contrast, for the nearly half-filled case [\figref{fig:mdf}{b}], the power-law features of $n(k)$ become indistinct for $V \geq 4t$, and the distribution evolves into a smooth function. While the system remains metallic, the Fermi point becomes ill-defined, indicating the breakdown of the LL description. At exact half-filling [\figref{fig:mdf}{c}], $n(k)$ exhibits no visible singularity for $V \geq 4t$, signaling a spontaneous symmetry-breaking transition into the CDW phase.

Moreover, we observe that, under fixed particle number, $n(k)$ always intersects the line $n(k) = n$ at $k_F = n \pi$, regardless of interaction strength. This feature persists across the LL, NLL, and CDW regimes, and may originate from a global sum-rule constraint imposed by particle-number conservation. Within the LL framework, such behavior can be attributed to the protection of the Fermi momentum. However, in other NFL regimes, a clear theoretical explanation remains elusive and may require a more detailed analysis of the Fourier structure of local density distributions. In any case, this observation suggests that certain features of the MDF are robust and insensitive to the specific nature of low-energy excitations or the details of critical behavior.
\begin{figure*}[htbp]
	\centering
	\includegraphics[width=1\linewidth,trim=25 20 19 15, clip]{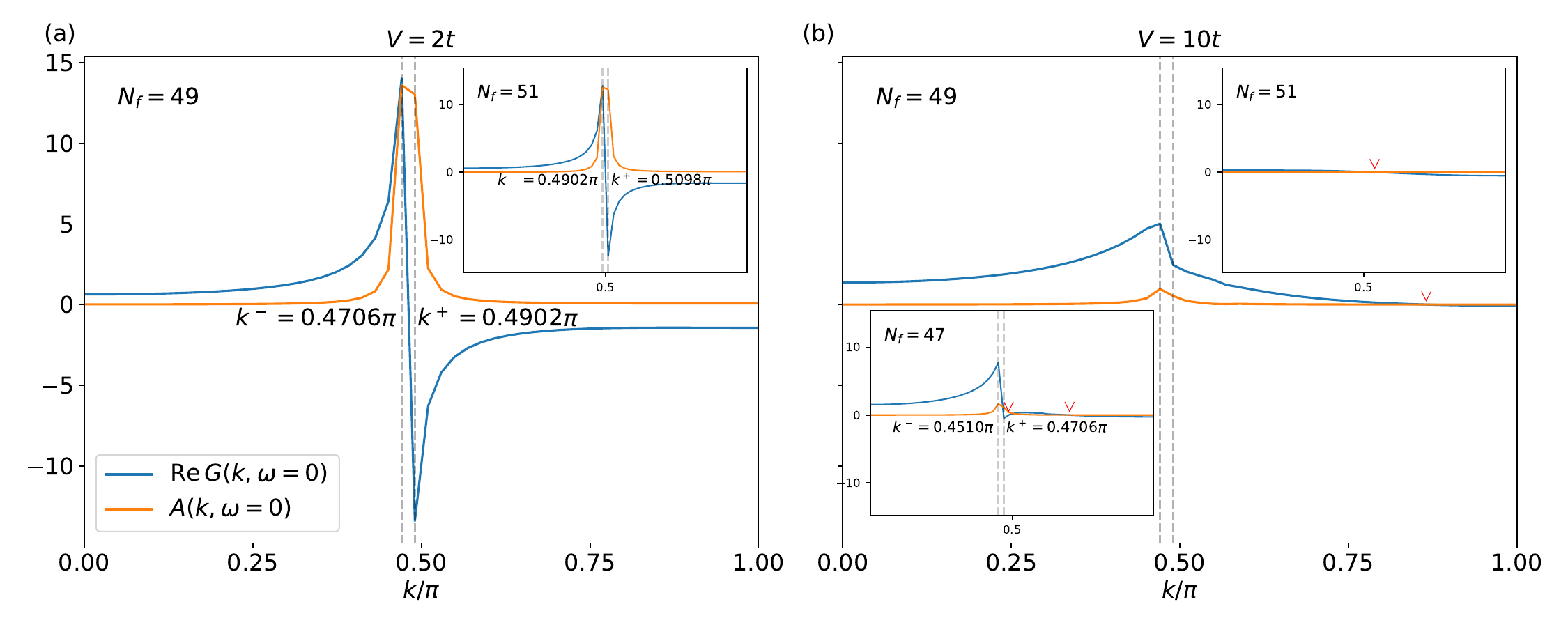}
	\caption{Green’s function $G(k, \omega = 0)$ and spectral function $A(k, 0)$ for a 102-site system under varying fillings and interactions. 
		The blue solid line shows the real part of the Green’s function \( \text{Re}\, G(k,0) \), and the orange solid line shows the corresponding \( A(k, 0) \). A damping factor of \( \eta = 0.05t \) is applied in all calculations. Both curves are rescaled and normalized to render them dimensionless.
		Dashed lines indicate the discrete momentum points closest to the Fermi momentum $k_F$, lying to its left and right, and denoted as $k^-$ and $k^+$, respectively. Red arrows indicate the locations of the zero crossings.
	}
	\label{fig:GF}
\end{figure*}

Based on numerical simulations for a 102-site system, \figref{fig:GF}{a} and \figref{fig:GF}{b} show the momentum-resolved $\mathrm{Re}\,G(k, 0)$ and the corresponding spectral function $ A(k, \omega) $ at two representative interaction strengths, $V / t = 2$ and 10, characterizing the typical features of the LL, NLL, and CDW phases. In \figref{fig:GF}{a}, the system exhibits LL behavior. In contrast, \figref{fig:GF}{b} shows results for particle numbers \( N_f = 47 \) and \( 49 \), corresponding to NLL states near half-filling, whereas the \( N_f = 51 \) panel represents the half-filled CDW phase.

In a finite-size lattice with discrete translational symmetry, the momentum values are quantized as
\begin{equation}
	k = \frac{2\pi l}{N}, \quad l = 0, 1, 2, \dots, N-1, 
\end{equation}
which in general do not exactly match the nominal Fermi momentum $k_F = (N_ f/N)\pi$. Nevertheless, low-energy features can still be captured effectively by analyzing the behavior around $k \approx k_F$ across the Brillouin zone.
\begin{figure*}[htbp]
	\centering
	\includegraphics[width=0.8\linewidth,trim=5 10 5 5, clip]{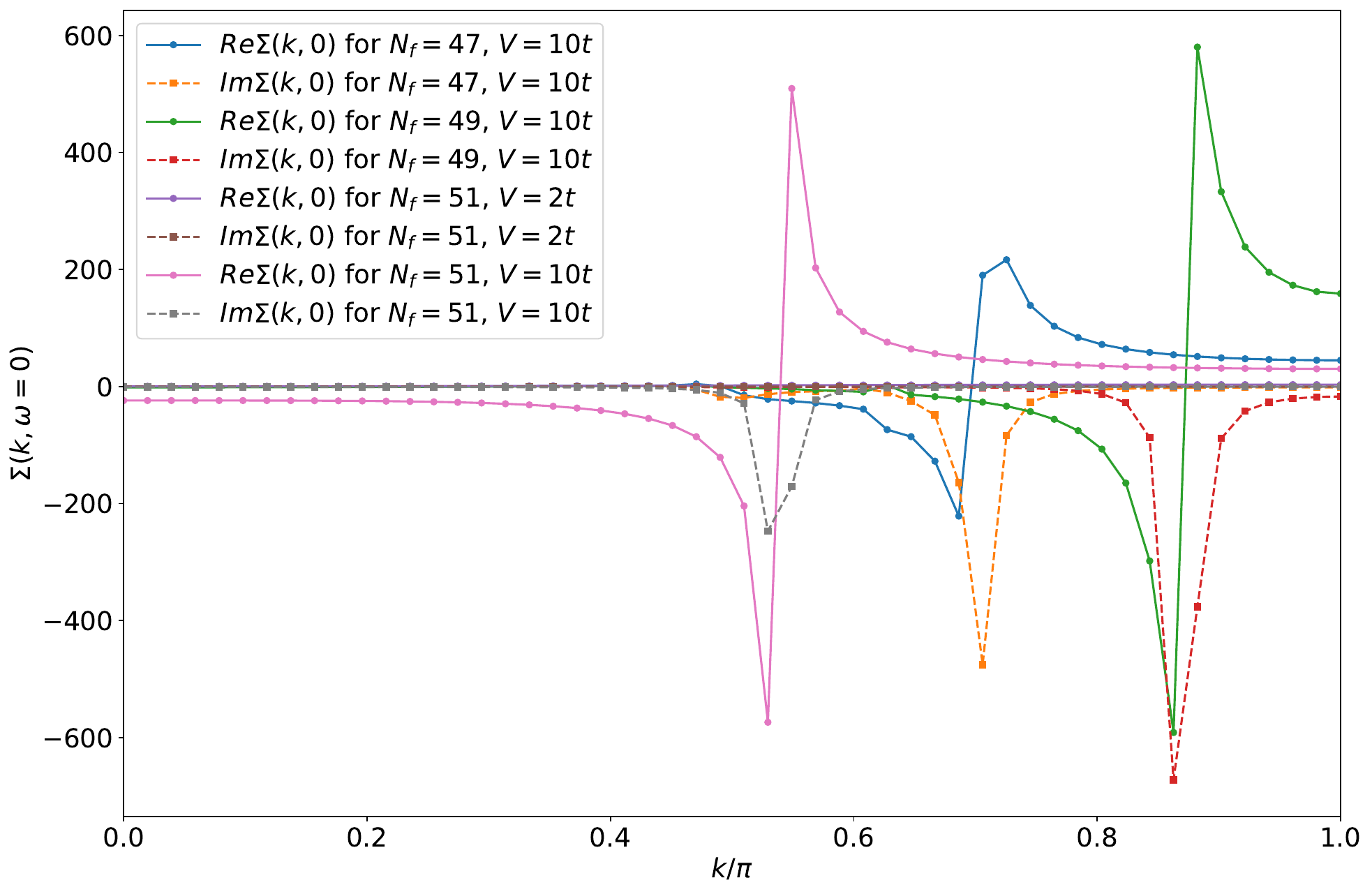}
	\caption{Self-energy $\Sigma(k, \omega = 0)$ for a 102-site system at various $N_f$ and $V$. In the LL phase, $\Sigma(k, 0) \approx \Sigma(0)$ remains a small finite value with negligible momentum dependence. In contrast, in the NLL and CDW phases, $\Sigma(k, 0)$ exhibits a rapid increase, or even divergence-like behavior, near the momentum points where $\mathrm{Re}\, G(k, 0) = 0$.  
	}
	\label{self-energy}
\end{figure*}

In the weak-coupling regime ($V < 2t$), the system resides in the LL phase. The low-energy behavior of $A(k,\omega)$ near the Fermi point is governed by the Luttinger parameter $K$. Under the linear dispersion approximation around $k_F$, $A(k,\omega)$ exhibits a characteristic power-law singularity~\cite{FDMHaldane_1981}, given by
\begin{equation}
 	A(k, \omega) \sim |\omega - v(k - k_F)|^{\alpha - 1}, 
\end{equation}
where \( v \) denotes the velocity of collective excitations near $k_F$. The exponent $\alpha$ is determined by $K$ via
\begin{equation}
 	\alpha = \frac{1}{2} \left( K + \frac{1}{K} - 2 \right). 
\end{equation}
For repulsive interactions, $K < 1$. The presence of next-nearest-neighbor hopping breaks p-h symmetry, resulting in an asymmetry between the spectral weights near $k^+$ and $k^-$ around $k_F$. As shown in \figref{fig:GF}{a}, the absence of a sharp quasiparticle peak and the broadening of $A(k, \omega=0)$ are characteristic of the LL phase. Unlike the FL, $\mathrm{Re}\,G(k, 0)$ no longer exhibits a pole-like (Lorentzian) singularity~\cite{PhysRevB.104.235122, PhysRevLett.110.090403} at $k_F$, but instead displays a power-law (branch-cut) singularity. Nevertheless, its sign structure remains similar to that of free fermions: $\mathrm{Re}\,G(k, 0) > 0$ for $k < k_F$ and $\mathrm{Re}\,G(k, 0) < 0$ for $k > k_F$. This sign change ensures the topological constraint required by the $I_L$, and implies that a generalized Fermi momentum can still be defined at the chemical potential.

As the filling approaches half-filling and $V/t$ increases, the system enters the NLL regime. As shown in \figref{fig:GF}{b} for $N_f = 47$ and $49$, the singularity in $\mathrm{Re}\, G(k, 0)$ is gradually smeared out, crossing zero smoothly at momentum points $k_L$, whose positions deviate increasingly from the weak-coupling $k_F$ as $V/t$ grows. Multiple such zero crossings may occur. $A(k, 0)$ becomes progressively smoother, though low-energy spectral weight persists, it is significantly suppressed compared to the LL regime, indicating a weakening of collective excitations.  
\begin{figure*}[htbp]
	\hspace*{-0.06\linewidth}  
	\centering
	\begin{tabular}{@{}cc@{}}
		\includegraphics[width=0.55\linewidth, trim=17 19 15 33, clip]{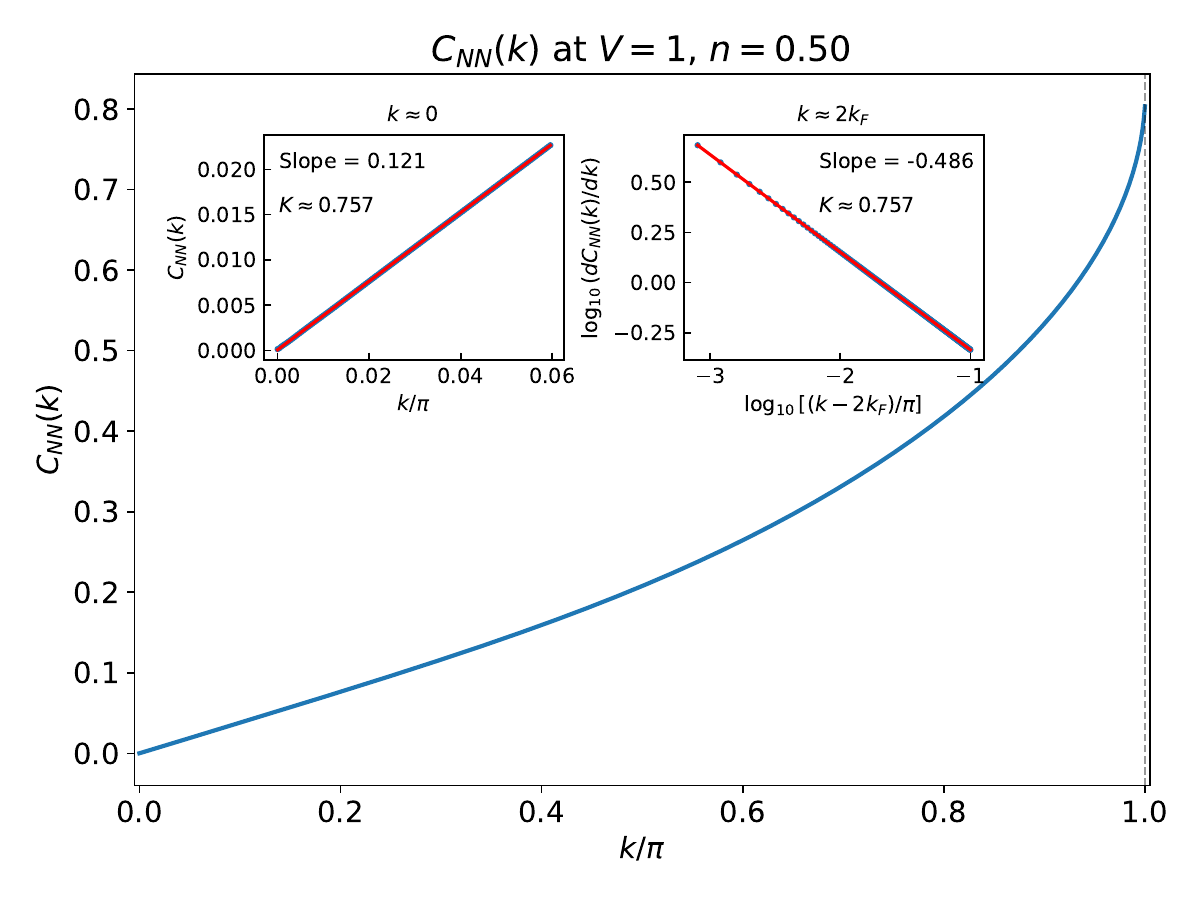} &
		\includegraphics[width=0.55\linewidth, trim=17 19 15 33, clip]{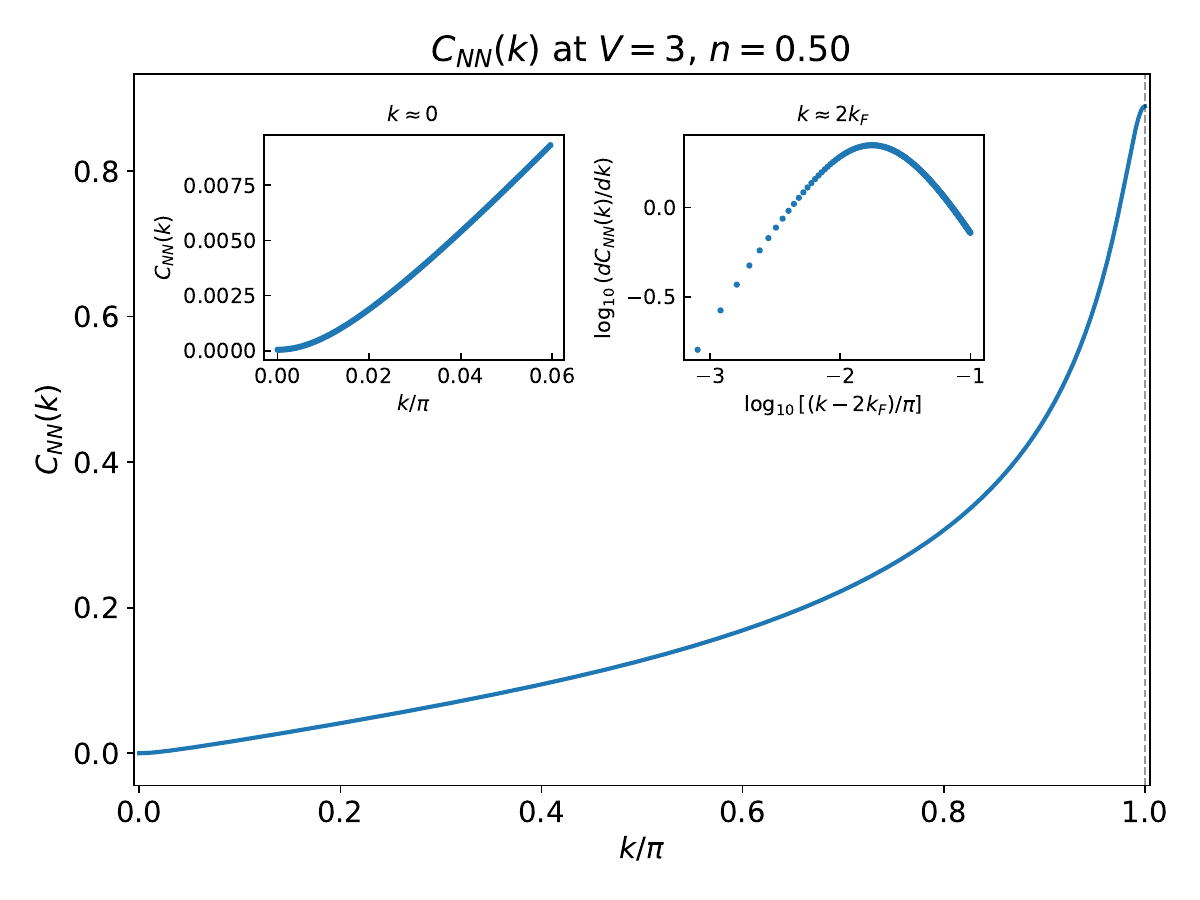} \\
		\small{(a) $V=t,n=0.50$} & \small{(b) $V=3t,n=0.50$} \\[2pt]
		
		\includegraphics[width=0.55\linewidth, trim=17 19 15 33, clip]{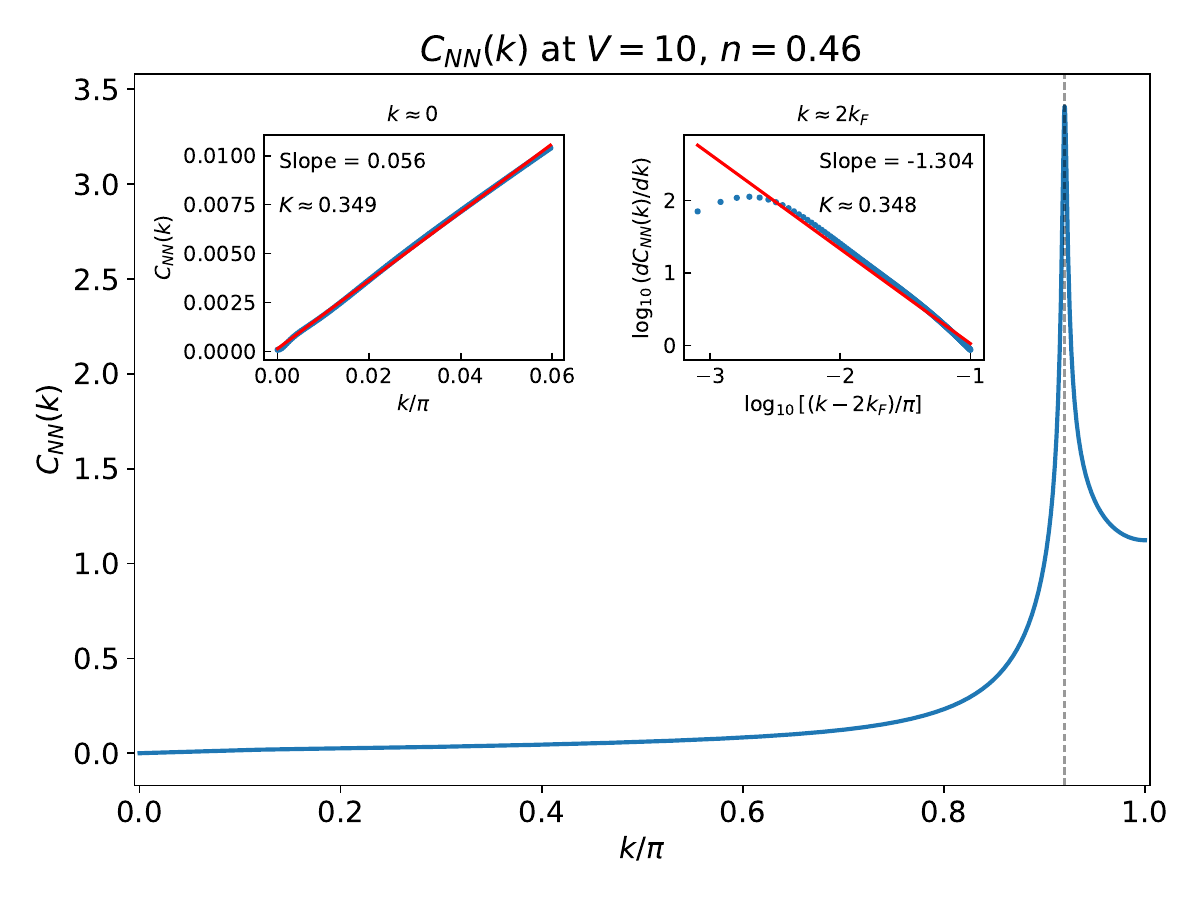} &
		\includegraphics[width=0.55\linewidth, trim=17 19 15 33, clip]{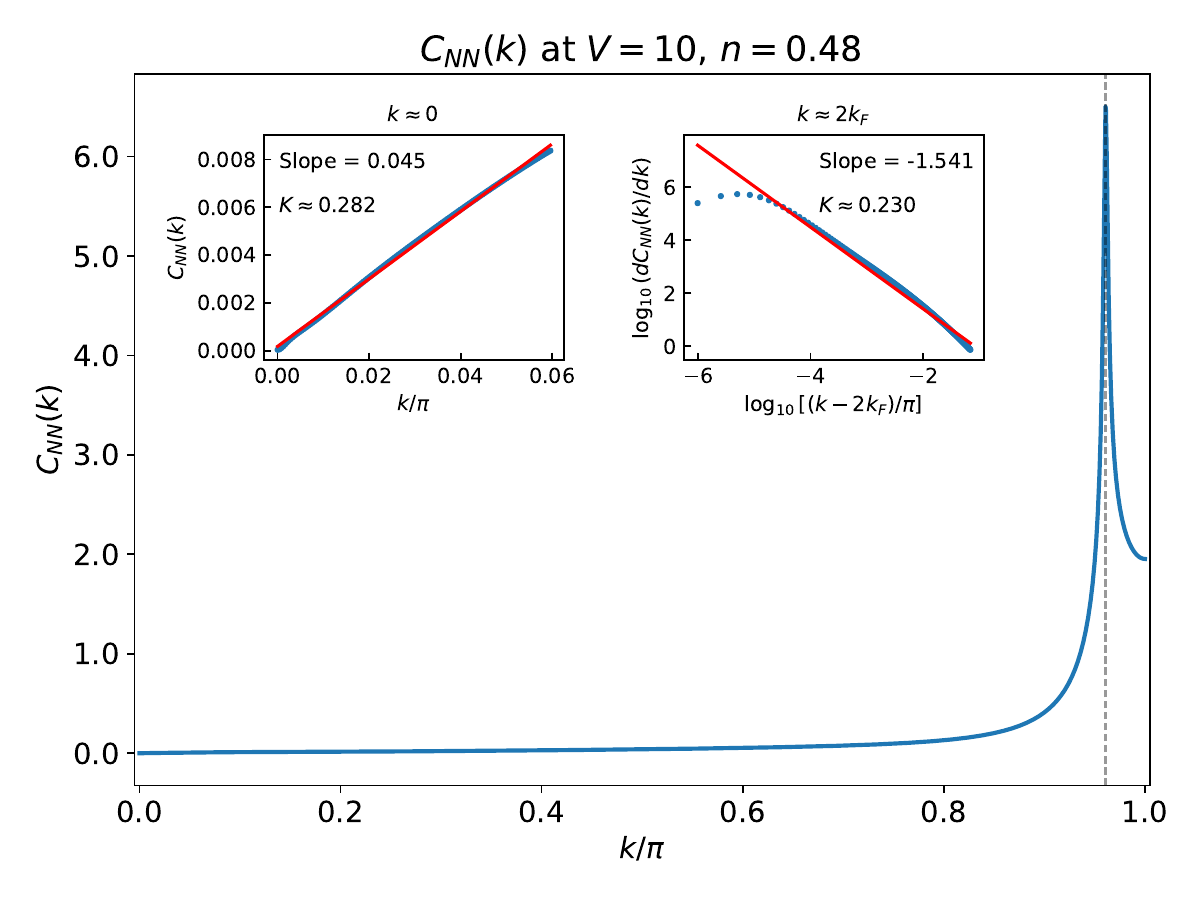} \\
		\small{(c) $V=10t,n=0.46$} & \small{(d) $V=10t,n=0.48$}
		
	\end{tabular}
	\caption{Momentum-space distribution of the density-density correlation function $C_{NN}(k)$ under various parameters near the thermodynamic limit (blue curve). In each subplot, the left and right insets show the derivative behavior of $C_{NN}(k)$ near $k = 0$ and $k = 2k_F$, respectively. Blue dots represent numerical data, while the red lines are linear fits in the corresponding regions, used to extract the Luttinger parameter and associated power-law features. Dashed lines mark the location of $2k_F$.}
	\label{fig:CNN}
\end{figure*}

With further enhancement of interaction strength, p-h asymmetry becomes more pronounced, and the self-energy $\Sigma(k, \omega=0)$ develops strong momentum dependence, as shown in Fig.~\ref{self-energy}. The divergence points of the self-energy align with the zero-crossings of $\mathrm{Re}\,G(k, 0)$ observed in \figref{fig:GF}{b}, a generic feature in both NLL and CDW regimes.

At half-filling, interactions open a charge gap, and single-particle excitations are fully suppressed in the CDW phase. As shown in the inset of \figref{fig:GF}{b} for $N_f = 51$, $G(k, 0)$ approaches zero across the entire momentum space. Within the interval $k \in [0, \pi]$, $\mathrm{Re}\,G(k,0)$ exhibits a single zero, whose deviation from $k_F$ is positively correlated with $V/t$. The peak of $A(k,0)$ completely vanishes, indicating the absence of accessible low-energy electronic states.
 
To distinguish the physical characteristics of the three phases above and to rule out artifacts arising from finite-size effects or numerical uncertainties, we compute the density-density correlation function of the model in the thermodynamic limit using the variational uniform matrix product state (VUMPS) algorithm:
\begin{equation}
	C_{NN}(k) = \frac{1}{N} \sum_{j,\ell} e^{ik(j - \ell)} \left( \langle \hat{n}_j \hat{n}_\ell \rangle - \langle \hat{n}_j \rangle \langle \hat{n}_\ell \rangle \right). 
\end{equation}
Results under different interaction strengths and fillings are compared in Fig.~\ref{fig:CNN}.

For the LL phase, the theoretical behavior of the derivative $\frac{dC_{NN}(k)}{dk}$ is expected to follow:
\begin{equation}
	\frac{d C_{NN}(k)}{dk} = 
	\begin{cases} 
		K/(2\pi), & k=0 \\[6pt] 
		\sim (2k_F - k)^{2K-2}, & k \approx 2k_F.
	\end{cases} 
\end{equation}

As shown in \figref{fig:CNN}{a}, by fitting the derivative near $k = 0$ and $k \approx 2k_F$, we observe power-law singularities consistent with LL theory~\cite{PhysRevB.86.155156}. In the critical region between LL and NLL [\figref{fig:CNN}{c}], although the nearly linear segment around $k \approx 2k_F$ can still be approximately fitted to a power-law form, the overall trend clearly deviates from linearity and begins to resemble the CDW-like behavior shown in \figref{fig:CNN}{b}. This indicates that the NLL does not possess the strict power-law character of an LL. Furthermore, as shown in \figref{fig:CNN}{d}, the LL signature becomes entirely obscured, suggesting that the NLL phase may involve a hybridization or competition between LL and CDW features, with the dominant behavior continuously evolving as the system parameters are tuned.

A closer comparison with $\mathrm{Re}\,G(k,0)$ in Fig.~\ref{fig:GF} reveals that as long as a singular structure is preserved near $k_F$, the correlation function $C_{NN}(k)$ typically retains a prominent feature at $2k_F$, reflecting the underlying Luttinger liquid-like behavior. However, in the NLL and CDW regimes, the zero-crossing behavior of $\mathrm{Re}\,G(k,0)$ often lacks the necessary singular support or disrupts the global topological structure of $G(k, 0)$ ~\cite{PhysRevLett.133.126504, NatCommun_14_4277} in momentum space, leading to a gradual suppression or smearing of the $2k_F$ peak.
\begin{figure*}[htbp]
	\centering
	\includegraphics[width=0.865\linewidth,trim=5 10 5 5,clip]{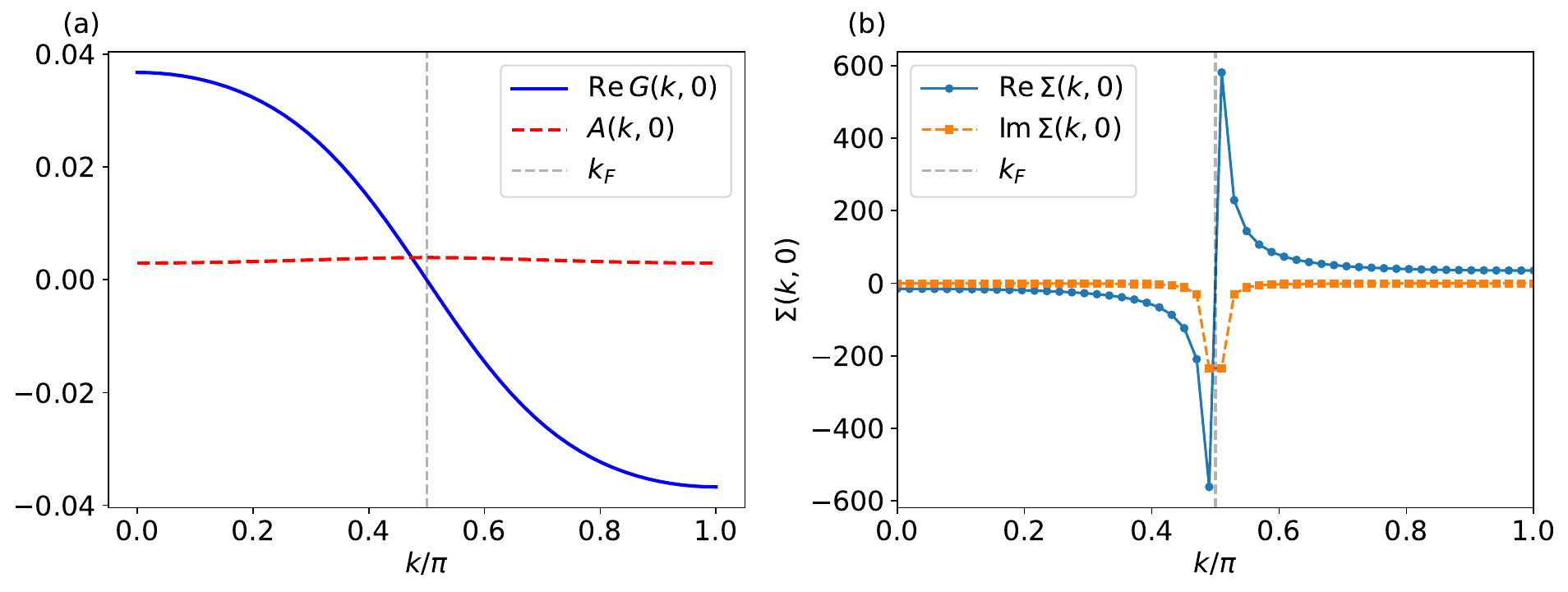}
	\caption{%
		$G(k,0)$ and $\Sigma(k,0)$ at half-filling ($n=0.5$), with $t'=0$ and strong interaction $V/t=10$. 
		(a) $\mathrm{Re}\,G(k,0)$ (blue solid), exhibits a zero crossing at $k=k_F$, while $A(k,0)$ (red dashed) vanishes at the same point due to the opening of a gap. 
		(b) The corresponding self-energy $\Sigma(k,0)$ shows a divergent real part and a peak in the imaginary part near $k_F$. 
		Both quantities remain symmetric about the Fermi momentum, illustrating that the Luttinger theorem remains valid despite the insulating CDW nature of the ground state.
	}
	\label{fig:tp_0}
\end{figure*}

Our analysis shows that the validity of the Luttinger theorem hinges on the zero-crossing behavior, broadly including both zeros and divergences of $\mathrm{Re}\,G(k, 0)$. On the one hand, divergences of $\mathrm{Re}\,G(k, 0)$ (e.g., poles or branch cuts) correspond to singular features protected by topology, thereby ensuring the theorem’s applicability. On the other hand, even if $\mathrm{Re}\,G(k, 0) = 0$ at certain momenta, the Luttinger theorem can still hold provided the relevant functions remain symmetric around $k_F$. This is exemplified by the half-filled system with $t' = 0$ and strong interaction $V/t = 10$, which forms a rigorously proven CDW insulator. Despite the insulating nature, the Luttinger theorem remains valid in this case, as evidenced in Fig.~\ref{fig:tp_0}, due to the symmetric structures of both $G(k, 0)$ and its self-energy around $k_F$. Thus, the presence of zero points does not necessarily alter the integral—the key lies in whether the Green’s function maintains an overall symmetry or hosts a topologically meaningful singular structure across the Fermi surface.

\section{Summary and Discussion}\label{Summary and Discussion}
In this work, we investigate the applicability of the Luttinger theorem in interacting fermionic systems using the 1D extended $t$-$V$ model. Our results demonstrate that in the LL phase, the \( I_L \) coincides with the particle density, indicating the robustness of the theorem in the weak-coupling regime. However, significant deviations arise in the strong-coupling regime, particularly in the CDW phase at half-filling and the NLL phase near half-filling.

We computed the zero-frequency Green's function $G(k, 0)$ for systems with 102 sites under various particle densities and interaction strengths. Importantly, finite-size effects do not qualitatively affect our conclusions, as verified by finite-size scaling, indicating that the observed features persist in the thermodynamic limit.

The redistribution of spectral weight is attributed to symmetry breaking and many-body reconstruction under strong coupling. In higher dimensions, weakly interacting systems can often be adiabatically connected to free fermions and described by FL theory. In contrast, in 1D, even infinitesimal interactions destabilize the free fermion point, driving the system toward an LL fixed point under renormalization group flow. Enhanced correlations eventually lead to NLL phases where $\mathrm{Re}\,G(k,0)$ becomes featureless, the singularity at the Fermi momentum fades, and the spectral weight near \( k_F\) is substantially suppressed. This reflects a global reconstruction of the spectral function and self-energy. In this regime, residual spectral features mainly signal nonperturbative collective excitations. At half-filling and strong coupling, the system opens a gap and enters the CDW phase, where low-energy single-particle excitations are completely suppressed.

The Luttinger theorem is known to hold in systems with p-h symmetry~\cite{PhysRevB.96.085124}. In the 1D $t$-$V$ model, introducing next-nearest-neighbor hopping $t'$ or deviating from half-filling explicitly breaks this symmetry, leading to asymmetric spectral weight around $k_F$. Interestingly, compared to the NLL regime, the CDW phase at half-filling exhibits partial symmetry restoration (in the interaction sector), which accounts for the relatively smaller deviation of \( I_L \) in this case.
\begin{figure*}[htbp]
	\centering
	\includegraphics[width=0.9\linewidth,trim=5 20 5 15, clip]{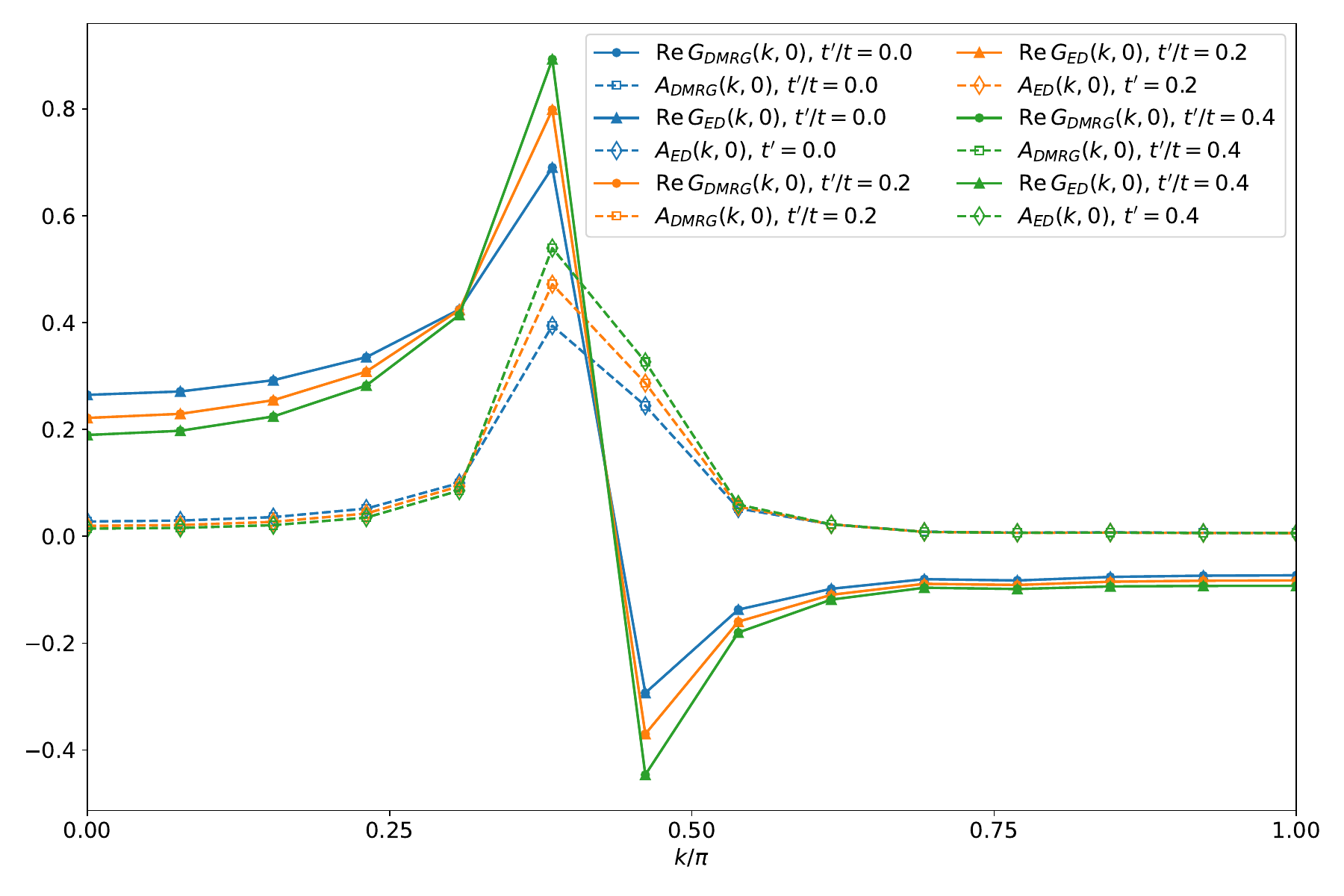}
	\caption{(Color online) Normalized $G(k, 0)$ computed using exact diagonalization and density matrix renormalization group for a system of size $N = 26$ at filling $N_f/N = 11/26$. Results are shown for hopping ratios $t'/t = 0$ (blue), 0.2 (orange), and 0.4 (green), with interaction strength $V/t = 7$.}
	\label{fig:ED_vs_DMRG}
\end{figure*}

A more precise description is that in the weak-coupling limit, LL and FL phases satisfy the Luttinger theorem due to emergent p-h symmetry. In the strong-coupling regime, $G(k,0)$ often develop zeros at certain momenta where $\operatorname{Re} G(k,0) = 0$, implying a divergent self-energy $\Sigma(k,0)$. In such cases, the Luttinger-Ward functional, which generates the free energy from $\Sigma$, becomes ill-defined. Nevertheless, this divergence does not necessarily invalidate the theorem. For systems with p-h symmetry, $G(k,0)$ and related quantities are often (anti)symmetric about $k_F$, thus preserving the total Fermi volume. Consequently, the presence of poles, branch cuts, or even nonsingular zeros does not affect the validity of the theorem as long as symmetry is maintained. In symmetry-broken systems, however, smooth zero crossings in $\operatorname{Re} G(k,0)$ can still lead to divergent self-energies, and more critically, the absence of topological constraints allows zero points to drift, ultimately invalidating the theorem. We propose that the interplay between the evolution of excitation modes and symmetry breaking breaks the original topological constraints, leading to a shift in the position of the generalized Fermi surface and its topological reconstruction, thereby driving the system away from the LL fixed point toward a new phase. This cooperative effect is not merely a simple superposition of the two mechanisms, but could represent a genuinely new type of collective phenomenon whose physical implications have not yet been systematically explored.

The behaviors of the Green’s function and self-energy described here are not unique to our system, similar features also appear in the Hatsugai-Kohmoto (HK) model~\cite{Zhao:2024puy}. In our case, although the system resides in a NLL state, the fermionic spectral weight remains finite near the original $k_F$, while it vanishes at the zero-defined $k_L$, accompanied by a pole-like self-energy that precludes a meaningful series expansion. What we aim to emphasize is that paradigmatic non-FL, such as LL~\cite{FDMHaldane_1981}, heavy-fermion metals~\cite{coleman2007heavy}, and strange metals~\cite{RevModPhys.73.797, RevModPhys.94.041002}, generally lack a unified low-energy expansion and a well-defined theoretical framework, and often originate from fundamentally distinct microscopic mechanisms. Nevertheless, they still exhibit certain commonalities, reflecting the intrinsic complexity and emergent phenomena of the non-FL. This diversity of exotic yet interconnected quantum states lies at the heart of strongly correlated physics and continues to motivate the search for a more universal and coherent theoretical framework.

In summary, our study shows that the breakdown of the Luttinger theorem in low-dimensional strongly correlated systems is not accidental, but a result of the interplay between many-body spectral reconstruction, symmetry breaking, and the evolution of excitation modes. By calculating the Luttinger integral, momentum distribution, and analyzing correlation functions, we propose a diagnostic framework for identifying and quantifying non-FL behavior. This work provides theoretical insights into the unconventional dynamics of low-dimensional systems, offering testable predictions for cold atom experiments, ARPES, and inelastic scattering probes~\cite{Yoshida_2012}. Extending this approach to more complex systems, such as the Hubbard model, $t$-$J$ model, and Kondo lattice~\cite{PhysRevLett.84.3370}, may shed further light on the general mechanisms underlying the failure of the Luttinger theorem and its experimental manifestations.

\begin{acknowledgments}	
This work was supported by the Supercomputing Center of Lanzhou University, which provided essential computational resources. We also acknowledge support from the National Natural Science Foundation of China (Grant No. 12247101), the Fundamental Research Funds for the Central Universities (Grant No. lzujbky-2024-jdzx06), the Natural Science Foundation of Gansu Province (Grant Nos. 22JR5RA389 and 25JRRA799), and the ‘111 Center’ under Grant No. B20063.
\end{acknowledgments}

\appendix*
\section*{Appendix: Benchmark Comparison Between DMRG and ED}\label{app:benchmark}

The DMRG method used in this work is a high-precision algorithm for ground-state simulations. The wavefunction is represented in the MPS form, and local optimizations combined with singular value decomposition (SVD) are applied to truncate the reduced density matrix, retaining only the most entangled states~\cite{SCHOLLWOCK201196}. This enables an efficient compression of the Hilbert space. While ED suffers from exponential scaling with system size, DMRG exploits the area law of entanglement entropy in 1D systems, reducing the computational cost to polynomial order (typically $O(\chi^3)$, with $\chi$ the MPS bond dimension). In practice, we systematically increase $\chi$ until the ground-state energy and relevant observables converge within the desired tolerance.

To verify the accuracy of our DMRG simulations, we benchmarked them against ED on small systems where the latter is feasible. Specifically, we considered the $t$-$t'$-$V$ model with PBC at various fermion densities, and computed the single-particle Green's function $G(k, 0)$ for $N_f/N = 11/26$, $V/t = 7$, and $t'/t = 0, 0.2, 0.4$. In the DMRG simulations, a truncation threshold of $10^{-10}$ was imposed, and small noise terms were introduced to avoid trapping in local minima. The sweeps were terminated once the energy change fell below $10^{-13}$, ensuring global convergence of the wavefunction and the physical reliability of the results.

Fig.~\ref{fig:ED_vs_DMRG} presents the results of $G(k, 0)$ obtained from DMRG and ED under the above parameters. As shown, excellent agreement is achieved throughout the Brillouin zone, with relative deviations below $10^{-6}$.
  
These benchmarks confirm that DMRG accurately captures the ground-state properties of the system, even for observables requiring high momentum resolution. Therefore, it is justified to apply DMRG to larger systems ($N=82$ to $102$) where ED becomes computationally intractable.

\bibliography{references}
\end{document}